\documentclass[12pt,psfig,showpacs,letterpaper]{revtex4}
\usepackage{geometry}
\usepackage{graphicx}
\usepackage{natbib}
\usepackage{amsmath}
\usepackage{amssymb}
\begin{document}
\bibliographystyle{acm}

\title{Creation of Magnetic Fields by Electrostatic and Thermal Fluctuations}
\author{Hamid Saleem\\
National Centre for Physics (NCP),\\
Quaid-i-Azam University Campus, Islamabad,\\ Pakistan.}
\date{September 01, 2009}

\begin{abstract}
It is pointed out that the unmagnetized inhomogeneous plasmas can
support a low frequency electromagnetic ion wave as a normal mode
like Alfven wave of magnetized plasmas. But this is a coupled mode
produced by the mixing of longitudinal and transverse components of
perturbed electric field due to density inhomogeneity. The ion
acoustic wave does not remain electrostatic in non-uniform plasmas.
On the other hand, a low frequency electrostatic wave can also exist
in the pure electron plasmas. But the magnetic field fluctuations in
both electron as well as in electron-ion plasmas are coupled with
the electrostatic perturbations in unmagnetized case. The main
instability condition for these low frequency electrostatic and
electromagnetic modes is the same $\frac{2}{3}\kappa_n < \kappa_T$
(where $\kappa_n$ and $\kappa_T$ are inverse of the scale lengths of
density and electron temperature, respectively).
\end{abstract}

\maketitle

\section{Introduction}
A low frequency electromagnetic wave has been investigated as a
fundamental normal mode of unmagnetized inhomogeneous plasmas. The
ion acoustic wave (IAW) is a well-known low frequency mode of
unmagnetized plasmas but it is purely electrostatic. Even for
lightest ions of hydrogen the electron to ion mass ratio is very
small $\frac{m_{e}}{m_{i}}\simeq10^{-3}<<1$. In the limit
$m_{e}\rightarrow0$, the electrons are assumed to follow the
Boltzmann density distribution in the electrostatic field of the
IAW. \\In fact the electron inertia can play a vital role in
producing a low frequency electromagnetic wave. More than a decade
ago \cite{saleem1}, it was proposed that an electromagnetic wave
having frequency near IAW is a normal mode of unmagnetized plasmas
which can be responsible for magnetic field generation if it becomes
unstable. In the derivation of its linear dispersion relation the
electron inertia was not ignored while the displacement current was
neglected. It was shown that the compressibility and vorticity can
couple due to density inhomogeneity and hence a low frequency
electromagnetic wave can be produced. Basically electrostatic IAW
and magnetostatic mode \cite{chu1} cooperate with each other to
develop such a wave. The electron temperature perturbation was not
taken into account and the steady state was assumed to be maintained
by external mechanisms. The theory was applied to explain the
magnetic field generation in laser plasmas. But the longitudinal and
transverse characters of electric field decouple if the
quasi-neutrality is used \cite{saleem1}. Then both the high
frequency and low frequency electromagnetic instabilities were also
investigated in unmagnetized plasmas \cite{saleem2}. \\The interest
in the investigation of low frequency magnetic fluctuations in
unmagnetized plasmas was initiated after the first experimental
observation of magnetic field generated in a laser induced plasma
\cite{stam1}. Then more experiments were performed \cite{stam2,
raven1} on these lines. Several theoretical models were presented to
explain the magnetic
field generation in laser produced plasmas \cite{altercop1, bol1, brueckner1, pert1, max1, haines1, jones1}.\\
The magnetic electron drift vortex (MEDV) mode was proposed as a
pure transverse linear mode which can exist because of the electron
temperature fluctuations in unmagnetized inhomogeneous electron
plasmas \cite{jones1}. This mode can become unstable \cite{yu1} if
the equilibrium electron temperature gradient is parallel to the
density gradient and it is maintained by external effects. The basic
MEDV mode is believed to exist in an electron plasma with smooth
gradients. However, it has also been shown that the instability of
MEDV mode can arise when the density profile is represented by a
single step connecting two regions of nonzero density and
temperature gradients \cite{yu2}. The instabilities of magnetic and
acoustic waves driven by perturbed baroclinic vector have also been
investigated in a pure electron plasma \cite{stenflo1}. Recently the
nonlinear evolution of two dimentsional MEDV modes has been studied
using computer simulation \cite{eliasson1}. The spontaneous magnetic
field generation and formation of nonlinear structures has been
discussed in detail. Most of the mechanisms proposed to explain
magnetic fluctuations in initially unmagnetized plasmas are based on
electron magnetohydrodynamics (EMHD) which has been discussed in
detail in Refs. \cite{king1, bol2}. Some weaknesses and
contradictions of EMHD model were pointed out several years ago
\cite{saleem3}. The long-lived and slowly propagating nonlinear
whistler structures (NLWS) or whistler spheromaks (WSPS) have also
been studied \cite{eliasson2} using EMHD equations. Such structures
have been observed in magnetized laboratory plasmas \cite{stenzel1,
stenzel2}.
\\In MEDV mode the divergence of electric field is assumed to be
zero $(\nabla.\textbf{E}_1)=0$ while the divergence of electron
velocity is non-zero $(\nabla.\textbf{v}_{e1}\neq0)$. Furthermore
the ions are treated to be stationary. The frequency $\omega$ of the
MEDV mode is assumed to lie in between the ion plasma frequency and
electron plasma frequency i.e. $\omega_{pi}<<\omega<<\omega_{pe}$
where $\omega_{pj}=\left(\frac{4\pi n_{0j}e^{2}}{m_{j}}\right)$, for
j=e,i and c is speed of light while k is the wave vector. These
restrictions and assumptions are indeed very strict \cite{saleem3}
and are
not fulfilled in general. \\
Several authors have considered the role of ion dynamics in the
magnetic instabilities in unmagnetized plasmas. The effects of ion
dynamics on MEDV mode have been investigated in the frame work of
local approximation \cite{mirza1}. The coupling of high frequency
electromagnetic wave with low frequency ion acoustic wave has been
discussed in certain limits \cite{saleem2}. The coupling of magnetic
fluctuations with ion acoustic wave has been discussed in a plasma
with the steady state given as $\nabla p_{e0}=0$ \cite{vranjes1}
without considering electron temperature perturbation. But the group
velocity
turns out to be negative in this treatment.\\
It is important to find out some electromagnetic mode taking into
account the ion dynamics using minimum approximations so that the
strict restrictions on the frequency and wavelength of the
perturbation are relaxed. It is better if the only required
condition on frequency becomes $\omega<<\omega_{pe}, ck$. If we
assume a steady state as $\nabla p_{e0}=0$, then the temperature
gradient becomes anti-parallel to density. But laser and
astrophysical plasmas  are open systems and many external mechanisms
can maintain a study state with parallel density and temperature
gradients. For example, in stellar cores, both the density and
temperature increase towards centre of the star due to fusion and
star is held intact because of gravity. Therefore both the cases of
parallel and anti-parallel gradients should be discussed. The
electron thermal fluctuations can produce electromagnetic wave as
was proposed many decades ago \cite{jones1} but the assumptions used in this work are very restrictive. \\
It is not necessary to assume a pure transverse perturbation in an
electron plasma. Rather the perturbation can be partially transverse
and partially longitudinal and hence the electron density
perturbation may not be neglected. The frequency of such a wave in
electron plasma turns out to be near $(\lambda_e
k_y)v_{te}\kappa_{n}$ where
$v_{te}=\left(\frac{T_{e}}{m_{e}}\right))^{\frac{1}{2}}$ is the
electron thermal speed, $\lambda_e =\frac{c}{\omega_{pe}}$ is the
electron skin depth and
$\kappa_{n}=|\frac{1}{n_{0}}\frac{dn_{0}}{dx}|$ is the inverse of
density gradient scale length $L_{n}=\frac{1}{\kappa_{n}}$. In the
local approximation we need to have $\kappa_{n}<<k$. Furthermore the
condition $\omega_{pi}^{2}<<v^{2}_{te}k^{2}$ can be satisfied if
$\frac{m_{e}}{m_{i}}<<\lambda^{2}_{De}k^{2}_{y}$. Therefore,
generally we may have $v_{te}\kappa_{n}\lesssim\omega_{pi}$ and
hence one cannot neglect ion dynamics. Moreover,
$v_{te}\kappa_{n}\simeq c_{s}k$ (where
$c_{s}=\left(\frac{T_{e}}{m_{i}}\right)^{\frac{1}{2}}$ is ion sound
speed) is also possible. Therefore, it is expected that the electron
thermal fluctuations can couple with IAW to produce stable and
unstable low frequency electromagnetic waves in unmagnetized
plasmas. It will be shown that the dispersion relation of ion
acoustic wave is modified in the inhomogeneous plasma due to the
coupling of electrostatic and magnetic fluctuations. A low frequency
electrostatic mode can do also exist in a non-uniform pure electron
plasma. This mode can couple with the IAW in electron-ion plasma.
Several low frequency electrostatic and electromagnetic waves of
inhomogeneous unmagnetized electron and electron-ion plasmas are
investigated. Interestingly the main instability condition for these
modes is the same.

\section{Low Frequency Electromagnetic Waves in Electron Plasmas}
Let us consider the electron plasma in the background of stationary
ions. First we discuss the dispersion relation of pure transverse
MEDV mode \cite{jones1, yu1}. Then we show that the compressibility
cannot be neglected. Finally linear dispersion relations of low
frequency electrostatic and electromagnetic perturbations are
obtained in an electron plasma. The set of equations for MEDV mode
in the linear limit can be written as,
$$m_e n_0
\partial_t \textbf{v}_{e1}=-en_0 \textbf{E}_1-\nabla
p_{e1}\eqno{(1)}$$$$\nabla\times\textbf{B}_1=\frac{4\pi}{c}\textbf{J}_1\eqno{(2)}$$$$\textbf{J}_1=-en_0
\textbf{v}_{e1}\eqno{(3)}$$$$\nabla\times\textbf{E}_1=-\frac{1}{c}\partial_t
\textbf{B}_1\eqno{(4)}$$ Since $p_1=n_0 T_{e1}$ therefore energy
equation becomes,
$$\frac{3}{2}n_{0}\partial_{t}T_{e1}+\frac{3}{2}n_0
(\textbf{v}_{e1}.\nabla)T_{e0}=-p_{0}\nabla.\textbf{v}_{e1}\eqno(5)$$
Curl of (1) gives,
$$\partial_{t}(\nabla\times\textbf{v}_{e1})=\frac{e}{m_{e}c}\partial_{t}\textbf{B}_{1}
+\frac{1}{m_{e}n_{o}}(\nabla n_0)\times\nabla T_{e1}\eqno(6)$$
Equations $(2)$ and $(3)$ yield, $$\textbf{v}_{e1}=-\frac{c}{4\pi
en_{0}}\nabla\times\textbf{B}_{1}\eqno(7)$$ and hence
$$\nabla\times\textbf{v}_{e1}=\frac{c}{4\pi
en_{0}}\nabla^{2}\textbf{B}_{1}\eqno(8)$$ where $\nabla
n_{0}\times(\nabla\times\textbf{B}_{1})=0$ due to the assumption
$\textbf{k}\bot\nabla n_{0}\bot\textbf{B}_{1}.$ Equation (8)
predicts $\textbf{E}_{1}=E_{1}\mathbf{\hat{x}}$ while $\nabla
n_{0}={\bf{\hat{x}}}\frac{dn_{0}}{dx}$, $\nabla=(0,ik_y,0)$ and
$\textbf{B}_1=B_1 \hat{\textbf{z}}$ have been chosen. \\
Equations (6) and (8) yield,
$$(1+\lambda^{2}_{e}k^{2})\partial_{t}\textbf{B}_{1}=-\frac{c}{en_{0}}(\nabla
n_{0}\times\nabla T_{e1})\eqno(9)$$ where
$\lambda_{e}=\frac{c}{\omega}_{pe}$. Equation (7) yields,
$$\nabla.\textbf{v}_{e1}=\frac{c}{4\pi n_{0}e}\frac{\nabla
n_{0}}{n_{0}}.(\nabla\times\textbf{B}_{1})\eqno(10)$$ and therefore
one obtains,
$$T_{e1}=\frac{2}{3}\frac{c}{4\pi n_{0}e}k_{y}\kappa_{n}B_{1}\eqno(11)$$
Then (9) and (11) give the linear dispersion relation of MEDV mode
as,
$$\omega^{2}=\frac{2}{3}C_{0}(\frac{\kappa_{n}}{k_{y}})^{2}v^{2}_{Te}k_{y}^2\eqno(12)$$
where $C_{0}=\frac{\lambda^{2}_{e}k^{2}}{1+\lambda^{2}_{e}k^{2}}$
and $v_{te}=(T_e/m_e)^{\frac{1}{2}}$. The geometry of MEDV mode in
cartesian co-ordinates is shown in Fig. 1.
\begin{figure}[htb]
        {\includegraphics[width=\textwidth]
        {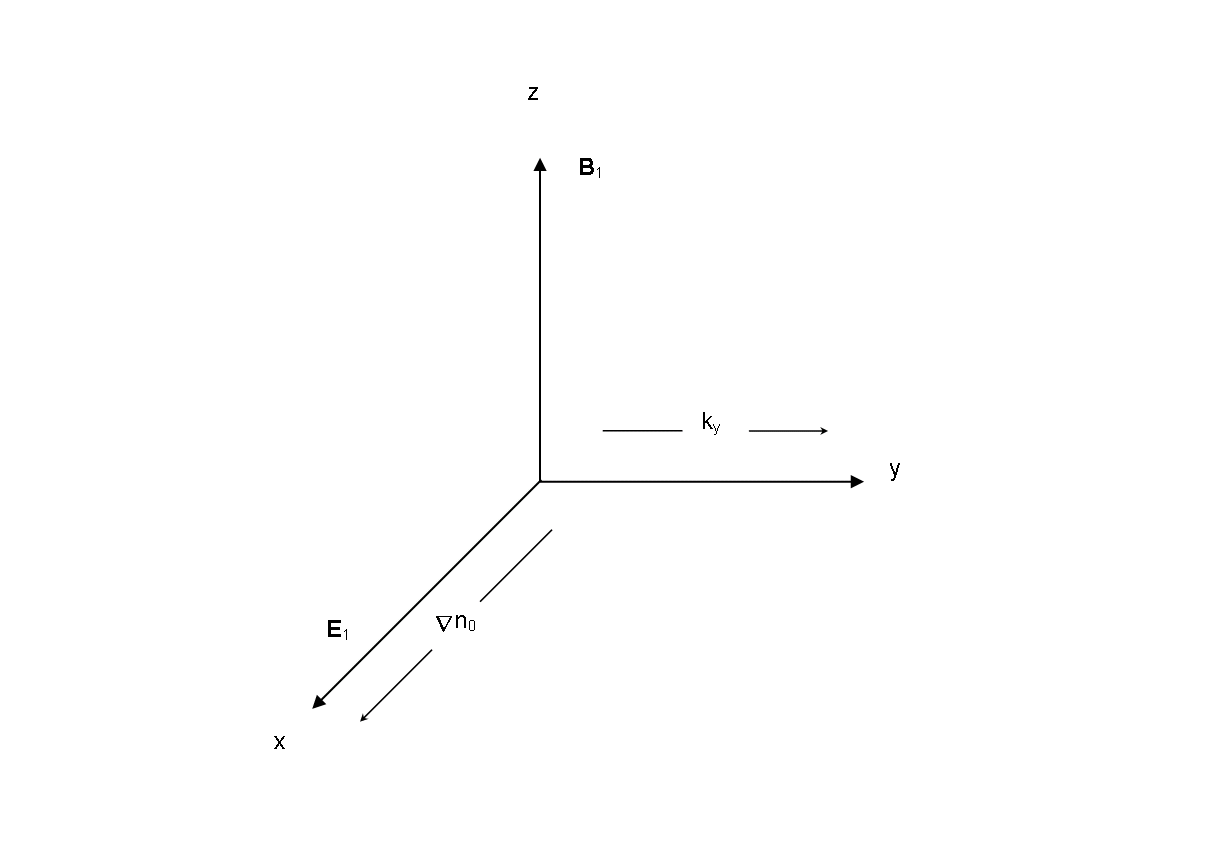}}
        \caption{The MEDV mode geometry shows that it is a pure transverse mode while the electron equation of motion indicates
        $E_{y1}\neq0$ and hence $\textbf{k}.\textbf{E}_{1}\neq0$.}
      \end{figure}\\ If $\nabla T_{e0}\neq0$
is assumed, then $(12)$ becomes,
$$\omega^{2}=C_{0}\frac{\kappa_{n}}{k_{y}}\left[\frac{(\frac{2}{3}\kappa_{n}
-\kappa_{T})}{k_{y}}\right]v^{2}_{Te}k^{2}\eqno(13)$$ where
$\kappa_{T}=|\frac{1}{T_{e0}}\frac{dT_{e0}}{dx}|$ and $\nabla
T_{e0}=+\textbf{{x}}\frac{dT_{e0}}{dx}$ has been used. If
$(\frac{2}{3}\kappa_{n}<\kappa_{T}),$ then the mode becomes unstable
\cite{yu1}.\\It can be noticed that $\nabla.\textbf{E}_1=0$ and
$\nabla.\textbf{v}_{e1}\neq0$ have been assumed in the above
treatment and it does not seem to be very convincing. If we write
(1) in x and y components, we can observe that the y-component of
$\textbf{E}_1$ should not be considered as zero due to $\partial_y
p_{e1}\neq0$ and hence the wave may be partially transverse and
partially longitudinal. Equation (1) yields,
$$\partial_{t}\nu_{ex1}=-\frac{e}{m_{e}}E_{x1}-v^{2}_{Te}\left\{\kappa_{n}\frac{T_{e1}}{T_{0}}+\kappa_{T}\frac{n_{e1}}{n_{0}}\right\}\eqno(14)\\$$
and
$$\partial_{t}\nu_{ey1}=-\frac{e}{m_{e}}E_{y1}-v^{2}_{Te}\left\{ik_{y}\left(\frac{T_{e1}}{T_{0}}+\frac{n_{e1}}{n_{0}}\right)\right\}\eqno(15)$$
It is obvious from (15), that $\nabla.\textbf{v}_{e1}\neq0$ which
implies $E_{y1}=-\partial_{y}\varphi_{1}\neq0$. It is clear from
Fig. 1 that MEDV mode should be revisited including longitudinal
effects.\\ The curl of (2) yields a relation between ${E}_{1x}$ and
${E}_{1y}$ for $\omega<<\omega_{pe}$ as,
$${E}_{1x}=-\frac{1}{a}\frac{\kappa_n}{k_y}(iE_{1y})\eqno{(16)}$$
where $a=(1+\lambda_{e}^{2} k_{y}^{2})$.\\Using equation of motion
(1) instead of (10) in equation (5), we find,
$$W_{0}^{2}\frac{T_{e1}}{T_0}=v_{te}^{2} k_{y}^{2}\left(1-\frac{3}{2}\frac{\kappa_{T}^{2}}{k_{y}^{2}}
-\Gamma_{0}^{2}\right)\frac{n_{e1}}{n_0}-\frac{e}{m_e}\left(ik_y
E_{1y}+\frac{3}{2}\kappa_{T} E_{1x}\right)\eqno{(17)}$$ where
$W_{0}^{2}=\frac{3}{2}\omega^2 -v_{te}^{2}
k_{y}^{2}\left(1-\frac{3}{2}\frac{\kappa_{T}
\kappa_n}{k_{y}^{2}}\right)$ and $\Gamma_{0}^{2}=\frac{(\kappa_T -
\kappa_n)\kappa_T}{k_{y}^{2}}$.
\\The continuity equation yields,
$$L_{0}^{2}\frac{n_{e1}}{n_0}=-\left(1+\frac{v_{te}^{2} k_{y}^{2}}{W_{0}^{2}}(1-\kappa_{n}^{2}/k_{y}^{2})\right)\left(i\frac{e}{m_e}k_y
E_{1y}\right)$$$$-\left\{1+\frac{3}{2}\frac{\kappa_T}{\kappa_n}\frac{v_{te}^{2}
k_{y}^{2}}{W_{0}^{2}}(1-\kappa_{n}^{2}/k_{y}^{2})\right\}\left(\frac{e}{m_e}\kappa_n
E_{1x}\right)\eqno{(18)}$$ where
$L_{0}^{2}=\left\{\omega^2-v_{te}^{2}k_{y}^{2}(1-\kappa_{n}^{2}/k_{y}^{2})-\frac{v_{te}^{4}k_{y}^{4}}{W_{0}^{2}}(1-\kappa_n^{2}/k_{y}^{2})\left(1-\frac{3}{2}
\frac{\kappa_{\Gamma}^{2}}{k_{y}^{2}}-\Gamma_{0}^{2}\right)\right\}$.
The Poisson equation
$$\nabla.\textbf{E}_1=-4\pi
e\left(\frac{n_{e1}}{n_0}\right)\eqno{(19)}$$ can be written as,
$$ik_y E_{1y}\left[L_{0}^{2}W_{0}^{2}-\omega_{pe}^{2}\left\{W_{0}^{2}+v_{te}^{2}k_{y}^{2}(1-\kappa_{n}^{2}/k_{y}^{2})\right\}\right]$$
$$=(\kappa_n E_{1x})\omega_{pe}^{2}\left[W_{0}^{2}+\frac{\kappa_T}{\kappa_n}v_{te}^{2}k_{y}^{2}(1-\kappa_{n}^{2}/k_{y}^{2})\right]\eqno{(20)}$$
\\Equations (16) and (20)
yield a linear dispersion relation in the limit $\omega^2 <<
\omega_{pe}^{2}$ as,
$$a\left[L_{0}^{2}W_{0}^{2}-\omega_{pe}^{2}\left\{W_{0}^{2}+v_{te}^{2}k_{y}^{2}\left(1-\kappa_n^2/k_{y}^{2}\right)\right\}\right]$$
$$=-\omega_{pe}^{2}(\kappa_n/k_{y})^2\left[W_{0}^{2}+\frac{3}{2}\frac{\kappa_T}{\kappa_n}v_{te}^{2}k_{y}^{2}\left(1-\kappa_n^2 / k_{y}^{2}\right)\right]\eqno{(21)}$$
This equation can be simplified as, $$H_0
W^2=-\frac{2}{3}v_{te}^{2}\kappa_{n}^{2}\left[\left(1-\frac{3}{2}\frac{\kappa_T}{\kappa_n}\right)-(1+\lambda_{e}^{2}k_{y}^{2})\left(1-\frac{3}{2}\frac{\kappa_T}{\kappa_n}\right)\right]\eqno{(22)}$$
where
$$H_0=\left[\left\{(1+\lambda_{De}^{2}k_{y}^{2})+\frac{2}{3}\lambda_{De}^{2}k_{y}^{2}\right\}a-\kappa_{n}^{2} / k_{y}^{2}\right]$$
The second term in right hand side of (22) will disappear if
$E_{1y}=0$ is assumed. The first term and the factor
$\lambda_{e}^{2}k_{y}^{2}$ in second term are the contributions of
transverse components $E_{1x}$. Thus one can not obtain MEDV-mode
dispersion relation for $E_{1y}=0$ from (22). It is interesting to
note that equation (22) can be simplified to obtain,
$$\omega^2=\frac{2}{3H_{0}}\lambda_{e}^{2}k_{y}^{2}(v_{te}^{2} \kappa_n)\left(1-\frac{3}{2}\kappa_T /
\kappa_n\right)\eqno{(23)}$$ which looks very similar to MEDV mode
dispersion relation.\\
 However, the instability condition for this
electromagnetic mode is the same as was for MEDV mode that is
$$\frac{2}{3}\kappa_n < \kappa_T\eqno{(24)}$$ The low frequency mode
(23) is partially transverse and partially longitudinal.\\
If $E_{1x}=0$ is assumed, then equation (21) yields a low frequency
electrostatic wave in a non-uniform unmagnetized plasma with the
dispersion relation,
$$\omega^2=\frac{v_{te}^{2}\kappa_{n}^{2}\left(\frac{2}{3}-\kappa_T / \kappa_n\right)}{\left[
(1+\lambda_{De}^{2}k_{y}^{2})+\frac{2}{3}\lambda_{De}^{2}k_{y}^{2}\right]}\eqno{(25)}$$
The instability condition remains the same (24). This indicates that
the instability predicted by the so called pure transverse MEDV mode
is always coupled with electrostatic perturbations in electron
plasmas.\\
In electron-ion plasmas, these modes of equations (23) and (25) can
couple with the ion acoustic wave.
\section{IAW and Magnetostatic Mode}
Here we shall show that transverse magnetostatic mode \cite{chu1}
which is obtained in the limit $\omega^2<<\omega_{pe}^{2}$ can
couple with IAW in a nonuniform plasma \cite{saleem1}. Therefore,
both ions and electrons are considered to be dynamic while the
electron temperature perturbation is neglected. The ions are assumed
to be cold for simplicity and therefore equation of motion becomes,
$$\partial_{t}\textbf{v}_{i1}=\frac{e}{m_{i}}\textbf{E}_{1}\eqno(26)$$ The continuity equation yields,
$$\frac{n_{i1}}{n_{0}}=\frac{e}{m_{i}\omega^{2}}(\kappa_{n}E_{1x}+ik_{y}E_{1y})\eqno(27)$$
If quasi-neutrality is used then the transverse component $E_{1x}$
and longitudinal component $E_{1y}$ become uncoupled \cite{saleem1}.
Therefore, we assume $\lambda^{2}_{De}k^{2}_{y}\neq0$ and use
Poisson equation which in the limit $\omega^{2}<<\omega^{2}_{pe}$
becomes,
$$\left[-v^{2}_{te}k^{2}_{y}\omega^{2}-\omega^{2}_{pi}(\omega^{2}-v^{2}_{te}k^{2}_{y})-\omega^{2}_{pe}\omega^{2}\right]ik_{y}E_{1y}\simeq[\omega^{2}_{pi}(\omega^{2}-v^{2}_{te}k^{2}_{y})+\omega^{2}_{pe}\omega^{2}]\kappa_{n}E_{1x}\eqno(28)$$
Note that the term $v^{2}_{te}k^{2}_{y}$ is not ignored compared to
$\omega^{2}_{pe}$ to couple $E_{1x}$ and $E_{1y}$.\\ Equations (16)
and (28) give a linear dispersion relation as,
$$\omega^{2}=\frac{c^{2}_{s}k^{2}_{y}(a-\frac{\kappa^{2}_{n}}{k^{2}_{y}})}{(ab-\frac{\kappa^{2}_{n}}{k^{2}_{y}})}\eqno(29)$$
where $c^{2}_{s}=\frac{T_{e}}{m_{i}}$, and
$b=(1+\lambda^{2}_{De}k^{2}_{y})$. We have to use Poisson equation
to obtain a quadratic equation in $\omega$ while (2) implies
$n_{e1}\simeq n_{i1}$. The equation (29) is the same as equation
(21) of Ref. \cite{saleem1} where two small terms in the denominator
are missing. Note that $\lambda^{2}_{De}<\lambda^{2}_{e}$ and if
quasi neutrality is used due to Ampere's law, then (29) yields the
basic electrostatic IAW dispersion relation
$\omega^{2}=c^{2}_{s}k^{2}_{y}$. If the displacement current is
retained and Poisson equation is used without using $\omega^{2},
\omega^{2}_{pi}<<\omega^{2}_{pe}$, then one obtains a dispersion
relation of coupled three waves; ion acoustic wave, electron plasma
wave and high frequency transverse wave \cite{saleem2}.
\\Actually the contribution of displacement current in the curl of
Maxwell's equation has been neglected for
$\omega^{2}<<\omega^{2}_{pe}, c^{2}k^{2}$. This should not mean that
the electrostatic part of current is also divergence free, in our
opinion. In the divergence part of Maxwell's equation
$\omega^{2}<<\omega^{2}_{pe}$ is used but $v^{2}_{te}k^{2}_{y}$ term
is assumed to be important. It may be mentioned that in this
treatment, Ampere's law does not imply quasi-neutrality necessarily.
We need a coupling of divergence part and curl part of the current
and for this we need to assume
$\frac{m_{e}}{m_{i}}<\lambda_{De}^{2}k_{y}^{2}$ in the limit
$\omega^2<<\omega_{pe}^{2}, c^2 k^2$.

\section{Low Frequency Electromagnetic Ion Waves}
 Now we present a simple but interesting theoretical model for
 low frequency electromagnetic waves assuming ions to be cold. The electrostatic waves will also
 be considered and it will be shown that magnetic field perturbation is coupled with the dominant
 electrostatic field. In the low frequency limit $|\partial_{t}|<<\omega_{pe}, ck$, the
 electrons are commonly assumed to be inertial-less, i.e.
 $\frac{m_{e}}{m_{i}}\rightarrow0$. Then the longitudinal and
 transverse components of electric field decouple. For IAW it is
 assumed that $E=-\nabla\varphi$ while electron equation of motion
 yields Boltzmann density distribution as
 $$\frac{n_{e}}{n_{0}}\simeq e^{-\frac{e\varphi}{T_{e}}}\eqno(30)$$
 Then the fundamental low frequency mode of the plasma turns out to
 be the ion acoustic wave with linear dispersion relation
 $$\omega^{2}_{s}=c^{2}_{s}k^{2}_{y}\eqno(31)$$ in the
 quasi-neutrality limit. If dispersion effects are included, then
 instead of (31) one obtains,
 $$\omega^{2}_{s}=\frac{c^{2}_{s}k^{2}_{y}}{1+\lambda^{2}_{De}k^{2}_{y}}\eqno(32)$$
 In the limit $1<<\lambda^{2}_{De}k^{2}_{y}$, equation (32) gives
 ion plasma oscillations $\omega^{2}=\omega^{2}_{pi}$. It is important to note that in the
 presence of inhomogeneity, a new scale $\frac{\kappa_{n}}{k_{y}}$
 is added to the system. If
 $\frac{m_{e}}{m_{i}}<<\left(\frac{\kappa_{n}}{k_{y}}\right)^{2}$,
 then longitudinal and transverse components of electric field can
 couple to generate low frequency electromagnetic waves.
 The divergence and curl of (1) give,respectively,
$$\partial_{t}\nabla.(n_{0}\textbf{v}_{e1})=-\frac{e}{m_{e}}n_{0}\nabla.\textbf{E}_{1}-\frac{e}{m_{e}}\nabla
n_{0}.\textbf{E}_{1}-\frac{1}{m_{e}}(\nabla.\nabla
p_{e1})\eqno(33)$$ and
$$\partial_{t}(\nabla\times\textbf{v}_{e1})+(\mathbf{\kappa}_{n}\times\partial_{t}\textbf{v}_{e1})=
-\frac{e}{m_{e}}\mathbf{\kappa}_{n}\times\textbf{E}_{1}-\frac{e}{m_{e}}\nabla\times\textbf{E}_{1}\eqno{(34)}$$
where $\kappa_{n}=|\frac{1}{n_{0}}\frac{dn_{0}}{dx}|$ and $\nabla
n_{0}=+\mathbf{\hat{x}}|\frac{dn_{0}}{dx}|$ has been assumed. If
initially electric field was purely electrostatic i.e.
$\textbf{E}_{1}=-\nabla\varphi_{1}$, then it will develop a rotating
part as well if $\nabla n_{0}\times\textbf{E}_{1}\neq0$, as is
indicated by the right hand side (RHS) of (34). \\ The Poisson
equation in this case is,
$$\nabla.\textbf{E}_1=4\pi n_0 e \left(\frac{n_{i1}}{n_0}-\frac{n_{e1}}{n_0}\right)\eqno{(35)}$$ Using (18) and (27),
the above equation can be written as,
$$
ik_{y}E_{1y}\left[L^{2}_{0}W^{2}_{0}\omega^{2}-L^{2}_{0}W^{2}_{0}\omega^{2}_{pi}-\omega^{2}_{pe}W^{2}_{0}+v^{2}_{te}k^{2}_{y}\left(1-\frac{\kappa_{n}^{2}}{k_{y}^{2}}\right)\right]
$$
$$
=\kappa_{n}E_{1x}\left[L^{2}_{0}W^{2}_{0}\omega^{2}_{pi}+\omega^{2}_{pe}\left\{W^{2}_{0}+
\frac{3}{2}\frac{\kappa_{T}}{\kappa_{n}}v^{2}_{te}k^{2}_{y}\left(1-\frac{\kappa_{n}^{2}}{k_{y}^{2}}\right)\right\}\right]\eqno{(36)}$$
Then (16) and (36) yield, $$a\left[L_{0}^{2}
W_{0}^{2}-\omega_{pe}^{2} (W_{0}^{2}+v_{te}^{2}
k_{y}^{2})\left(1-\kappa_n^2/k_{y}^{2}\right)\right]$$
$$=-\frac{\kappa_{n}^{2}}{k_{y}^{2}}\left[\omega_{pe}^{2}\left\{W_{0}^{2}+\frac{3}{2}\frac{\kappa_T}{\kappa_n}v_{te}^{2}\left(1-\kappa_n^2/k_{y}^{2}\right)\right\}\right]$$
$$+\frac{L_{0}^{2} W_{0}^{2}}{\omega^2}\omega_{pi}^{2}\left(a-\frac{\kappa_{n}^{2}}{k_{y}^{2}}\right)\eqno{(37)}$$
In the limit $\omega^2, \omega_{pi}^{2}<< \omega_{pe}^{2}$, (37)
gives a linear dispersion relation,
$$\omega^2=\frac{1}{H_0}\left[(\lambda_{e}^{2} k_{y}^{2}) v_{te}^{2}
\kappa_{n}^{2}
\left(\frac{2}{3}-\frac{\kappa_T}{\kappa_n}\right)+\left(a-\kappa_n^2/k_{y}^{2}\right)c_{s}^{2}k_{y}^{2}
\left\{\frac{5}{3}-\left(\frac{\kappa_T^2}{k_{y}^{2}}+\frac{\kappa_T
\kappa_n}{k_{y}^{2}}\right)\right\}\right]\eqno{(38)}$$ If ion
dynamics is ignored then (38) reduces to (23). The above equation
shows a coupling of ion acoustic wave with the electromagnetic
fluctuations in nonuniform unmagnetized plasmas.\\
In the electrostatic limit $(E_{1x}=0)$, equation (38) becomes,
$$\omega^2=\frac{1}{H_1}\left[v_{te}^{2}\kappa_{n}^{2} \left(\frac{2}{3}-\kappa_T/\kappa_n\right)+c_{s}^{2}k_{y}^{2}
\left\{\frac{5}{3}+\left(\frac{\kappa_T^2}{k_{y}^{2}}+\frac{\kappa_T
\kappa_n}{k_{y}^{2}}\right)\right\}\right]\eqno{(39)}$$ where
$H_1=\left\{(1+\lambda_{De}^{2}k_{y}^{2})+\frac{2}{3}\lambda_{De}^{2}k_{y}^{2}\right\}$. For stationary ions, (39) reduces to (25).\\
It is important to note that the main instability conditions for
electrostatic ion acoustic wave (39) and low frequency
electromagnetic wave (38) is again the same (24).\\
In electron-ion plasma, the electromagnetic wave dispersion relation
in the quasi-neutrality limit can be written as,
$$\omega^2=\frac{\lambda_{e}^{2}k_{y}^{2}}{(a-\kappa_{n}^{2}/k_{y}^{2})}v_{te}^{2}\kappa_n^2\left(\frac{2}{3}-\frac{\kappa_T}{\kappa_n}\right)
+\frac{5}{2}c_{s}^{2}k_{y}^{2}\eqno{(40)}$$ The wave geometry is
shown in Fig. 2. \\
The instability can occur if (24) is satisfied
along with
$$\frac{5}{2}c_{s}^{2}k_{y}^{2}< \frac{\lambda_{e}^{2}k_{y}^{2}}{(a-\kappa_{n}^{2}/k_{y}^{2})}v_{te}^{2}\kappa_n^2\eqno{(41)}$$
\begin{figure}[htb]
        {\includegraphics[width=\textwidth]
        {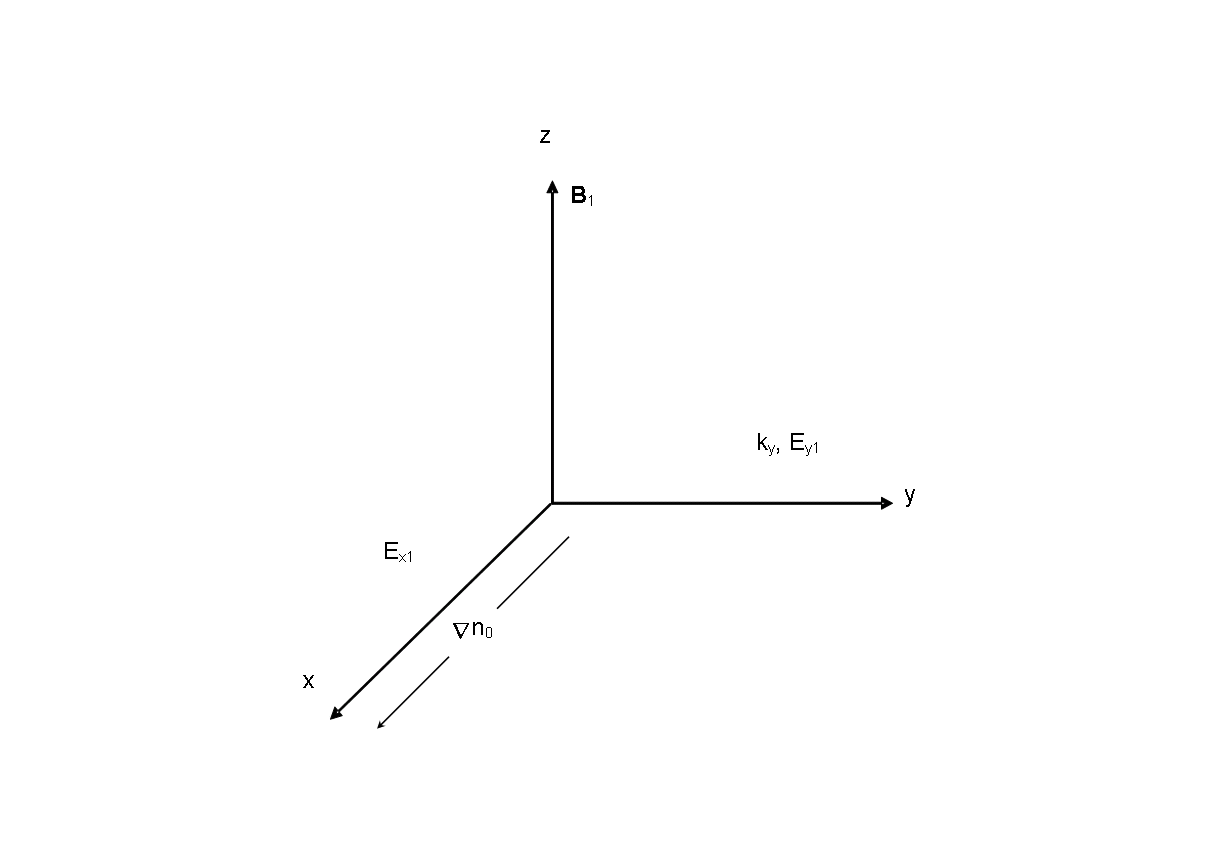}}
        \caption{The simplest possible geometry of electromagnetic ion wave is shown.
        This low frequency wave is partially longitudinal and partially transverse.}
      \end{figure}\\
\section{Discussion}
The theoretical model presented here shows that several
electrostatic and electromagnetic low frequency waves can exist in
un-magnetized electron as well as electron-ion plasmas.
Interestingly the main instability condition for these modes is the
same $\frac{2}{3}\kappa_n<\kappa_T$ where $\kappa_n$ and $\kappa_T$
are the inverse of density and electron temperature gradient scale
lengths, respectively. This indicates that the magnetic field
fluctuations are always coupled with the electrostatic
perturbations.\\
It is well-known that the ion acoustic wave is a fundamental low
frequency electrostatic mode of un-magnetized plasmas. Here we have
found that the inhomogeneous electron plasma can also support a low
frequency electrostatic mode. This electrostatic mode can couple
with the magnetic field perturbations to give rise to a partially
electrostatic and partially transverse wave. Therefore instead of
the so called magnetic electron drift vortex (MEDV) mode there
exists a low frequency electromagnetic wave having both the
contributions of longitudinal and transverse electric field
components as has been shown in equations (22). The dispersion
relation of MEDV mode is very similar to the electromagnetic wave
discussed in this investigation as can be seen in equation (23). But
the important point to note is that this dispersion relation appears
after a cancellation of two terms. One of these terms is a part of
longitudinal electric field $E_{1y}$. Moreover, if transverse
electric field component $E_{1x}$ is neglected one obtains a pure
electrostatic wave of equation (25). The frequency range of both the
modes is very close to each other. The instability conditions are
almost the same. This fact strengths the view point that magnetic
fluctuations are coupled with the dominant electrostatic fields.\\
Any initial electrostatic field perturbation can produce its
transverse component in the presence of density gradient
\cite{saleem1, saleem2}. This phenomenon can cause a coupling of ion
acoustic wave (IAW) with the low frequency transverse magnetostatic
mode. This coupled mode has already been investigated more than a
decade ago \cite{saleem1}. But it can exist in a relatively shorter
wavelength range for $\frac{m_e}{m_i}<\lambda_{De}^{2}k^2$. In the
quasi-neutrality limit, the IAW and magnetostatic modes decouple.\\
In case of the electron-ion plasmas, the IAW does not remain
electrostatic in inhomogeneous plasmas \cite{saleem1, vranjes1,
saleem4}. The electromagnetic mode discussed for the case of pure
electron plasma can couple with ion acoustic wave as shown in
equation (40). Similar to the electrostatic IAW, this low frequency
electromagnetic ion wave can exist even in the quasi-neutrality
limit. The electrostatic and electromagnetic waves discussed here
can become unstable if the density and temperature gradients are
parallel to each other which
can be the case in laser plasmas similar to stellar cores.\\
These low frequency waves can be the intrinsic source of magnetic
fields in stars, galaxies as well as in laser plasmas. The main
instability condition of the several electrostatic and
electromagnetic waves discussed here is the same as given in the
form of inequality (24). Therefore, in our opinion, electrostatic
perturbations are strongly coupled with magnetic field fluctuations
in inhomogeneous un-magnetized plasmas.

\pagebreak

\end{document}